\documentclass[twocolumn]{aastex701}

\usepackage{amsmath}
\usepackage{xcolor}
\usepackage{units}

\begin{document}

\title{Stochastic Variability of Binary Accretion}

\author{Akhil Nair}
\affiliation{Department of Physics and Astronomy, Clemson University, Clemson, SC 29634, USA}
\email[show]{akhiln@clemson.edu}

\author{Jonathan Zrake}
\affiliation{Department of Physics and Astronomy, Clemson University, Clemson, SC 29634, USA}
\email{jzrake@clemson.edu}

\begin{abstract}
  We measure the power spectral density (PSD) of the accretion rate time
  series in an unequal mass ($q=0.2$) binary surrounded by a circumbinary gas
  disk, using very high-resolution 2D hydrodynamics simulations. Our aim is to
  identify new signposts of supermassive black hole (SMBH) binaries in active
  galactic nuclei (AGN), based on the shape of the \emph{continuum} PSD, to
  complement well-studied \emph{line} features in the PSD (periodicities). We
  find that the continuum PSD is a broken power-law, transitioning from flat
  (white noise) to a slope of $-4$ at a break frequency generically $\sim 5$
  times the binary orbital frequency. This form is expected when (a) delivery
  of gas from the circumbinary disk to the individual ``minidisks'' is a damped
  random walk with correlation time equal to binary orbital period and (b) the
  minidisks function as low-pass filters acting at the Kepler frequency of the
  outer edge of the smaller black hole's minidisk; we show numerical evidence
  for both. The broken power-law PSD is attained in a limit where the secondary
  black hole is much smaller than its minidisk, realized numerically by a
  sufficiently small ``sink'' region; larger sinks lead to excess
  high-frequency noise seen as accretion rate spikes, and we argue these should
  be regarded as artificial when the black holes themselves are smaller than
  the sink regions. The broken power-law PSD is reminiscent of stochastic
  variability in ordinary AGN, inviting the conjecture that canonical AGN
  variability could result from widespread binarity, however pulsar timing
  experiments may exclude this possibility.
  \end{abstract}



\section{Introduction}

Supermassive black hole (SMBH) binaries are an expected consequence of hierarchical galaxy formation. When two massive galaxies merge, gravitational torques channel gas and stars toward the nuclear region \citep{MihosHernquist1996}, and the central black holes sink toward a common center through dynamical friction and stellar scattering, eventually forming a gravitationally bound pair \citep{BBR1980,MilosavljevicMerritt2001}. These systems are expected to be common at sub-parsec separations, and their study stands to significantly increase understanding of cosmic structure formation, the environments of galactic centers, and the production of low-frequency gravitational waves now evidenced by pulsar timing arrays \citep{NANOGrav2023}. Yet they remain observationally elusive. Sub-parsec binaries cannot be spatially resolved with current instruments, and the most widely pursued temporal signpost, quasi-periodic flux modulation at or near the binary orbital frequency \citep{Charisi2016,Liao2021,Chen2024}, is complicated by the intrinsically stochastic nature of AGN light curves \citep{Matthews1963,Ulrich1997,Padovani2017}. Apparent periodicity must be carefully distinguished from red noise fluctuations \citep{Vaughan2016,Witt2022}, and periodicity searches are limited to orbital periods sufficiently shorter than the survey baseline, typically of order a decade, which excludes the vast majority of the expected sub-parsec population \citep[e.g.][]{Haiman2009}, whose orbital periods extend to $\unit[10^{3}]{yr}$. Moreover, even for binaries within the accessible period range, periodogram recovery rates drop dramatically for the non-sinusoidal (e.g.\ sawtooth) pulse shapes predicted by hydrodynamic simulations \citep{Lin2026}.

These limitations motivate theoretical exploration of temporal signposts of binary SMBH pairs, beyond periodic brightness variations of the host nucleus. Gas accretion in binary systems is inherently variable and intermittent, including on timescales much shorter than the binary orbital period. This raises the possibility that binary AGN could exhibit stochastic variability signatures with a distinct character from that of single AGN.

In this paper we use hydrodynamics simulations to quantify the stochastic variability of black hole mass accretion rates arising uniquely from the chaotic circumbinary gas flows. We focus on sub-orbital timescales, aiming to identify a signpost of long-period SMBH binaries that could be revealed in high-cadence optical surveys such as the Vera C.\ Rubin Observatory LSST \citep{Ivezic2019}. Specifically, we measure the overall amplitude and shape of the \emph{continuum}
\footnote{By ``continuum'' we mean the part of the PSD aside from periodicities (which would then be analogous to ``lines'' in an emission spectrum). Throughout this paper ``accretion rate'' refers to the joint time series $\dot M_{1,2}$ of both the primary and secondary components.}
part of the accretion rate power spectral density (PSD): does it exhibit power-law scaling in frequency? Are there characteristic break frequencies? What is the rms variability on timescales much shorter than the orbital period? We hope to reveal ways in which the continuum PSD differs from the canonical red-noise spectrum of AGN \citep{Lyubarskii1997, Kelly2009}.

The literature contains numerous studies based on simulations of binary accretion to characterize quasi-periodic features of the accretion rate time series. For nearly equal mass circular binaries, the dominant low-frequency modulation is the ``lump'', an $m=1$ overdensity at the inner edge of the circumbinary disk that orbits at $\sim 0.2\,\Omega_{\rm bin}$ \citep{MacFadyenMilosavljevic2008,Shi2012,Farris2014,Dorazio2016,Duffell2020,Dittmann2022}. Power at the orbital frequency and its harmonics is also present, varying in amplitude with mass ratio. Quasi-periodic variability has also been seen in 3D GRMHD simulations associated with ``sloshing'' mass exchange between the minidisks at frequencies near the binary orbital period \citep{Avara2024}, and 2D simulations show that the minidisks themselves can develop eccentricity and collide with one another at roughly the orbital period \citep{WesternacherSchneider2024}. In eccentric prograde binaries with $e \gtrsim 0.1$, the dominant modulation occurs at the binary orbital frequency \citep{Zrake2021,DOrazio2021,WesternacherSchneider2022,Dorazio2024}.

In contrast, the \emph{continuum} of the PSD remains largely uncharacterized, particularly at super-orbital frequencies, thus novel measurements can be made within a simple physical setup. For our simulations in this first study on the subject, we assume the disk is thin and the dynamics are predominantly coplanar. We use a locally isothermal equation of state with orbital Mach number $\mathcal{M}=10$. This comes at the expense of being unable to compute emission spectra; we thus use accretion rates onto the component black holes as proxies for bolometric luminosity, which we believe is reasonable given that in a thin, radiatively efficient disk the thermal emission is expected to roughly track $\dot M$, especially at photon energies at and above optical-UV \citep[e.g.][]{WesternacherSchneider2022}. Disk internal stresses, nominally associated with unresolved MHD turbulence, are modeled using the Shakura-Sunyaev $\alpha$ prescription ($\alpha = 0.01$). We exclude magnetohydrodynamics (MHD), isolating the variability signal induced by binary accretion from variability associated with temporal fluctuations of the MRI stresses \citep{Lyubarskii1997}.

In this paper we also systematically explore the role of \emph{sink size}, a parameter that determines the inner truncation radii of the minidisks. Sink size can be loosely interpreted as the radii of the black holes' innermost stable circular orbits (ISCOs). For a binary system where the separation $a$ is much greater than the gravitational radius $r_g \equiv GM_{\rm bin}/c^2$ (with $M_{\rm bin}$ denoting the binary mass throughout), computational feasibility may require the numerical sink size to be significantly larger than the physical ISCO size, in which case the sink size must be regarded as a numerical parameter; its choice might thus influence measurements such as the component accretion rate time series. We characterize the sinks by a single dimensionless sink-size parameter $s$, defined such that the component sink radii are $r_{\rm sink,1} = s\,a\,(M_1/M_{\rm bin})$ and $r_{\rm sink,2} = s\,a\,(M_2/M_{\rm bin})$. We systematically vary $s$ in the range $0.05$ to $0.40$, and establish a criterion for $s$ to be small enough that the PSD converges to a universal shape, insensitive to the size of the sink.

We have selected a circular binary with a modest mass ratio, $q \equiv M_2/M_1 = 0.2$, for which $r_{\rm sink,1} = 0.833\,s\,a$ and $r_{\rm sink,2} = 0.167\,s\,a$. Varying $s$ then changes the ratio of the outer to inner minidisk radii, and thus the capacity of minidisks to ``buffer'' the gas supply from the CBD. We measure the rate of mass flowing through the secondary minidisk by integrating the mass flux through circular control surfaces surrounding the secondary BH. This allows us to separately examine the time series of mass entering the secondary disk, versus falling onto the secondary BH.

Our paper is structured as follows. Sec.~\ref{sec:numerical_setup} describes the simulation method, mesh geometry and motion, and sink implementation. Sec.~\ref{sec:results} presents the accretion rate PSD measurements and the cutoff frequency determination, and includes supporting diagnostics of the secondary minidisk inflow variability (Sec.~\ref{sec:minidisk_inflow}). Sec.~\ref{sec:minidisk_analytic} develops an analytic driver-filter model for the minidisk response. Sec.~\ref{sec:discussion} discusses implications for simulation methodology and for identifying sub-parsec SMBH binaries from continuum variability in AGN surveys. Sec.~\ref{sec:summary} summarizes our main conclusions.

\section{Numerical setup}
\label{sec:numerical_setup}
\subsection{Code Overview}
We solve the vertically-integrated compressible Navier–Stokes equations using the \texttt{Sailfish} code \citep{Zrake2024}. The code uses an explicitly conservative, second-order numerical scheme, and is written specifically for the numerical study of disk-binary interactions.

We adopt the locally isothermal equation of state, in which the sound speed in the disk is explicitly set to
\begin{equation}
  c_s(\vec r, t)=\frac{\sqrt{-\phi(\vec r, t)}}{\mathcal{M}}
\label{eq:cs}
\end{equation}

where the binary gravitational potential is
\begin{equation}
  \phi(\vec r, t)
    = -\frac{G M_1}{\sqrt{|\vec r - \vec x_1(t)|^2 + \epsilon^2}}
      -\frac{G M_2}{\sqrt{|\vec r - \vec x_2(t)|^2 + \epsilon^2}},
  \label{eq:phi}
\end{equation}
with $\vec x_{1,2}(t)$ the inertial-frame positions of the binary components and $\epsilon$, a Plummer softening length set to one cell width at $\tilde r = a$, and $\mathcal{M} \equiv 10$ is a globally defined orbital Mach number.
Gravity is included by treating the binary components as Newtonian point particles. Each binary component is surrounded by a ``sink'' region wherein gas is subtracted, details are provided in Sec. \ref{sec:sinks}.

\subsection{Mesh geometry and motion}
Our simulations use a polar mesh $(\tilde r, \phi)$ centered on the instantaneous position of the primary black hole $\vec x_1(t)$, logarithmic in $\tilde r$ and uniform in $\phi$, with $\tilde r \in [\,r_{\rm inner},\;10^3 a]$ and $\phi \in [0, 2\pi]$. The grid extends well inside the primary sink region. Zero-gradient outflow conditions are applied at the inner and outer radial boundaries.

The mesh center tracks the instantaneous position of the primary black hole
$\vec x_1(t)$ rather than the system barycenter. Because the grid spacing is
logarithmic, spatial resolution scales as $\tilde r\,\Delta\ln\tilde r$ and
is finest near the mesh origin. Centering on $M_1$ therefore places the
highest resolution around the primary sink, which for $q < 1$ has the larger
radius ($r_{\rm sink,1} > r_{\rm sink,2}$) and encloses the more massive
minidisk. The secondary orbits at $\tilde r = a$. For a circular binary, $M_1$ traces a circle of radius
$r_1 = q\,a/(1+q)$ about the center of mass ($r_1 \approx 0.167\,a$ for
$q = 0.2$), so the inertial-frame position of a grid vertex at mesh
coordinates $(\tilde r, \phi)$ is
\begin{equation}
  \vec x(\tilde r, \phi, t)
    = \tilde r\,(\cos\phi,\,\sin\phi) + \vec x_1(t).
  \label{eq:mesh_wobble}
\end{equation}
The mesh therefore ``wobbles'' at the orbital frequency, a new feature of \texttt{Sailfish} \citep{Zrake2024} first used in this work. Every cell moves with the common velocity $\dot{\vec x}_1(t)$. Intercell Godunov fluxes are evaluated across moving faces, with each face velocity including the $\dot{\vec x}_1(t)$ contribution, so the conservative update remains exact. The equations of motion are solved in the inertial frame, so no Coriolis or centrifugal source terms appear in the momentum equation. Gravitational source terms are evaluated at inertial-frame positions via Equation~\eqref{eq:mesh_wobble}.

We use a multi-resolution restart sequence: the simulation is first run at relatively low resolution to reach a statistically steady state, then upsampled by a factor of two for the measurement phase, and upsampled once more for a convergence check.

The mesh cell size is set to
\[
\Delta \ln \tilde r = \left\{
\begin{array}{ll}
0.0100 & \text{Phase I: burn-in, 2900 orbits},\\
0.0050 & \text{Phase II: sampling, 100 orbits},\\
0.0025 & \text{Phase III: convergence test, 100 orbits}.
\end{array}
\right.
\]
The 2900-orbit Phase~I duration corresponds to roughly three viscous times at $r = a$ for $\alpha = 0.01$. In Phase~II we force the code to use a fixed timestep so the component accretion rate time series are evenly sampled, which is more convenient for spectral analysis. Samples are recorded roughly $10^3$ times per binary orbit, enabling measurement of the PSD up to frequencies much higher than the binary orbital frequency.

\subsection{Viscosity model}
Viscous stresses in the disk are modelled using the standard \(\alpha\)–prescription \citep{ShakuraSunyaev1973}, in which the kinematic shear viscosity is set to
\begin{equation}
    \nu(\vec r, t) = \alpha \frac{c_s(\vec r, t)^2}{\tilde \Omega(\vec r, t)}
\label{eq:nu}
\end{equation}
where $\tilde \Omega(\vec r, t)^2 \equiv G M_1 / r_1^3 + G M_2 / r_2^3$ and $r_{1,2}$ are the distances between $\vec r$ and the respective black holes. We adopt \(\alpha = 0.01\) in Equation~\ref{eq:nu}.

\subsection{Initial conditions}
We initialize an axisymmetric locally isothermal disk, with the surface density $\Sigma(r)$, radial velocity $v_r(r)$, and azimuthal velocity $v_\phi(r)$ following the analytic viscous-disk solution \citep{Pringle1981} for a constant $\alpha$ prescription,
\begin{align}
  \Sigma(r)  &= \frac{\dot M}{3\pi\nu(r)} \\
  v_r(r)     &= -\frac{\dot M}{2\pi\,r\,\Sigma(r)} \\
  v_\phi(r)  &= \sqrt{\frac{GM_{\rm bin}}{r}\left(1 - \frac{1}{\mathcal{M}^2}\right)}.
  \end{align}
  Here $\nu(r)$ is the axisymmetric form of Equation~\ref{eq:nu} with $\Omega(r) = \sqrt{GM_{\rm bin}/r^3}$.

\subsection{Sink prescription}
\label{sec:sinks}
\begin{figure*}[!htbp]
  \centering
  \includegraphics[width=\textwidth]{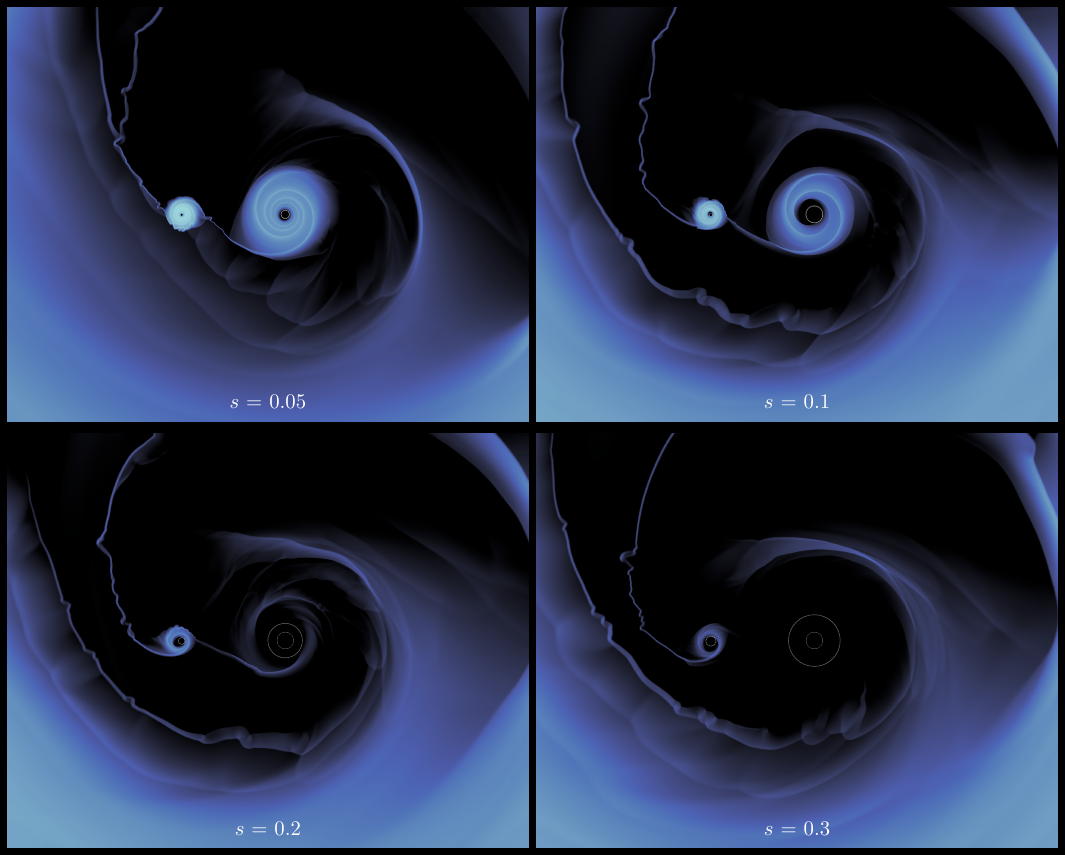}
  \caption{Two-dimensional surface density snapshots at t = 3100 orbits with different $s$ parameter values (0.05, 0.1, 0.2 and 0.3). Minidisks evolve from deep, strongly buffering structures at small $s$ values (0.05, 0.1) to progressively shallower structures at larger $s$ values (0.2, 0.3). The white circles indicate the sink boundaries.}
  \label{fig:massdensitycollage}
\end{figure*}

In a binary system with mass ratio $q = 0.2$, the component masses are
\[
M_1 = 0.8333\, M_{\rm bin},\quad M_2 = 0.1667\, M_{\rm bin}.
\]
Each black hole is surrounded by a sink region of radius
\[
R_i = s\, a\, \frac{M_i}{M_{\rm bin}} \,,\quad s \in \{0.05,\,0.10,\,0.20,\,0.30,\,0.40\},
\]
where $a$ is the binary semi-major axis.
Within \(r<R_i\), gas is removed at a rate
\begin{equation}
  \dot{\Sigma}_{\rm sink}(r) = -\tau_{\rm sink}^{-1}\, \Sigma(r) \, \exp\!\left[-\left(\frac{r}{R_i}\right)^4\right]
  \end{equation}
  where $\tau_{\rm sink}^{-1} = 100$ in code units. Fig.~\ref{fig:massdensitycollage} shows 2D surface density snapshots at $t = 3100$ orbits for four $s$ values. Minidisks evolve from deep, strongly buffering structures at small $s$ (0.05, 0.10) to progressively shallower structures at larger $s$ (0.20, 0.30). The largest case ($s = 0.40$, not shown) produces the shallowest minidisks.

\subsection{Convergence tests}

Convergence tests at $\Delta\ln r = 0.0025$ reproduce the same PSD features observed in the $\Delta\ln r = 0.005$ runs, confirming convergence in the measurement band.

\section{Results}
\label{sec:results}

We compute the power spectral density (PSD) of each sink accretion rate time series $\dot{M}_{1,2}(t)$. In the limit of small $s$ the PSD converges to a broken power-law shape, which we will show emerges from a two-stage stochastic process, referred to henceforth as a ``driver-filter'' process. The circumbinary streams supply variable gas to the outer edge of each minidisk (the driver), and the minidisk filters this supply before it reaches the sink (the filter).

\subsection{Measurement of the PSD}
\label{sec:results_methods}

We analyze the final 100 orbits of Phases~II and III for each accretion rate time series $\dot{M}_i(t_n)$, $n = 0, \ldots, N-1$, sampled at a uniform interval, yielding $\sim 10^3$ samples per binary orbit. The main PSDs shown below use the Phase~III interval at $\Delta\ln r = 0.0025$; the Phase~II interval is used as a lower-resolution comparison for convergence. The total duration $T = N\,\Delta t$ sets the frequency resolution $\Delta f = 1/T$ and the Nyquist limit $f_{\text{Nyq}} = (2\Delta t)^{-1}$. To reduce spectral leakage, we apply a Hann window, which tapers the signal smoothly to zero at the boundaries. The discrete Fourier transform of the windowed series is
\begin{equation}
X_k = \sum_{n=0}^{N-1} w_n\, \dot{M}_i(t_n)\, \exp(-2\pi i k n/N), \qquad 0 \leq k < N,
\end{equation}
and we retain the one-sided spectrum for $0 \le k \le N/2$
with frequencies
\begin{equation}
    f_k = \frac{k}{N\Delta t}.
\end{equation}
Because time is measured in binary orbits, $f_k$ is measured in
${\rm orbit}^{-1}$, and the binary orbital frequency corresponds to
$f=1$.  When comparing PSD breaks to local orbital motion, we quote
Keplerian frequencies in the same units,
\begin{equation}
    f_{K,i}(r) \equiv \frac{\Omega_{K,i}(r)}{\Omega_{\rm bin}},
    \qquad
    \Omega_{K,i}(r) = \left(\frac{G M_i}{r^3}\right)^{1/2},
\end{equation}
where $i=1,2$ labels the binary component.

The power spectral density is
\begin{equation}
P_i(f_k) = \frac{2\,\Delta t}{\sum_{n=0}^{N-1} w_n^2}\, \big|X_k\big|^2 .
\end{equation}

\begin{figure}
  \centering
  \includegraphics{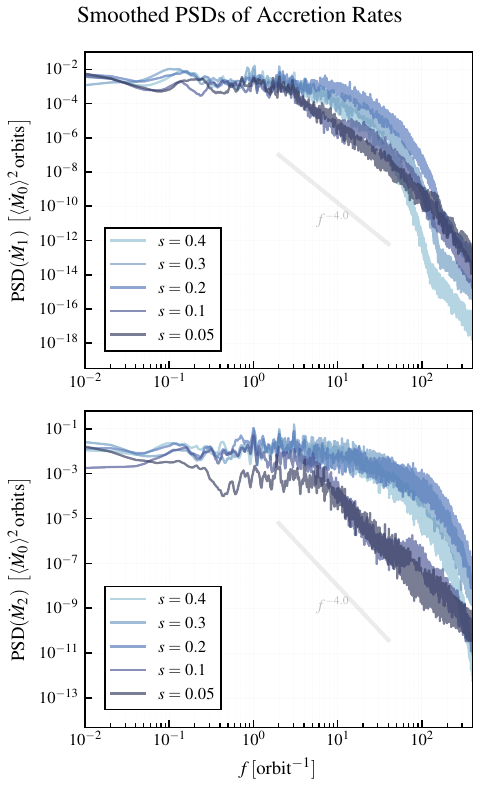}
  \caption{Smoothed power spectral densities of $\dot M$ for various $s$ values, sampled from the $\Delta\ln r = 0.0025$ run over 100 orbits (3000--3100).}
  \label{fig:psddx0025}
\end{figure}

\begin{figure}
  \centering
  \includegraphics{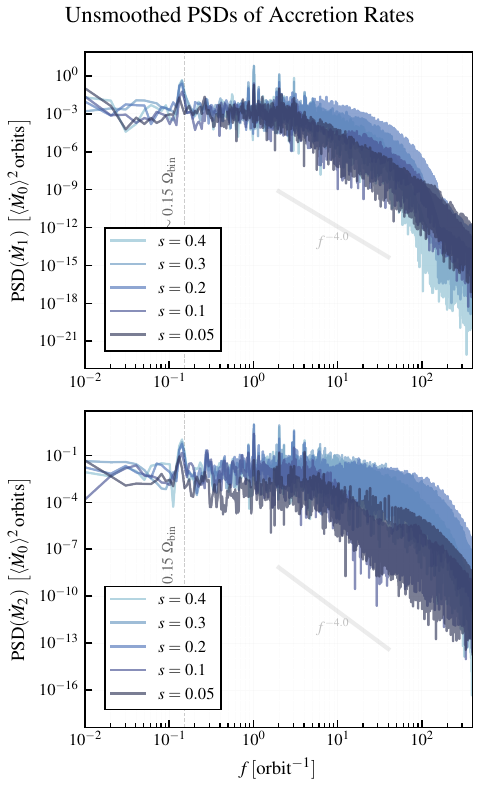}
  \caption{Unsmoothed power spectral densities of $\dot M$ for various $s$ values, sampled from the $\Delta\ln r = 0.0025$ run over 100 orbits (3000--3100). The dashed vertical line marks the ``lump'' frequency $f \approx 0.15$ in binary-orbit units, associated with the $m=1$ overdensity at the inner edge of the circumbinary cavity \citep{MacFadyenMilosavljevic2008}.}
  \label{fig:psddx0025_unsmoothed}
\end{figure}

Frequencies are plotted in units of orbit$^{-1}$, so the binary orbital frequency corresponds to $f = 1$. The PSD curves in Fig.~\ref{fig:psddx0025} are smoothed with a Savitzky-Golay filter \citep{SavitzkyGolay1964} for visual clarity. All quantitative measurements of break frequencies and spectral slopes are taken from the unsmoothed PSDs in Fig.~\ref{fig:psddx0025_unsmoothed}. Accretion rates are expressed in units of $\langle\dot{M}_0\rangle$, the steady-state mass supply rate through the circumbinary disk, and PSD amplitudes carry units of $\langle\dot{M}_0\rangle^2 \, \mathrm{orbit}$.

\subsection{Dependence on the sink size $s$}
\label{sec:morphology}

Figs.~\ref{fig:psddx0025} and \ref{fig:psddx0025_unsmoothed} show the smoothed and unsmoothed sink PSDs for $s \in \{0.05, 0.10, 0.20, 0.30, 0.40\}$. The main result is the small-sink limit. In this limit the PSD is consistent with an unresolved doubly broken continuum: a low-frequency plateau, a narrow orbital-to-outer-minidisk $f^{-2}$ band, and a steep tail close to $f^{-4}$ beyond $f_d \simeq 5$. At $q = 0.2$ the expected $f^{-2}$ band spans less than a decade, so the plateau and the $f^{-4}$ tail meet in a single rounded transition rather than a clean resolved segment. This is the converged small-sink shape, with the minidisk acting as a low-pass filter on the stream supply.

Large sinks do not produce a new converged shape. They distort the small-sink continuum by letting excess high-frequency power reach the sink, which broadens the high-frequency turnover. As $s \to 0$ this excess power is removed, and the rounded turnover settles to a clean power law with index $-4$, recovering the small-sink continuum.

\begin{figure*}[t]
  \centering
  \includegraphics[width=\textwidth]{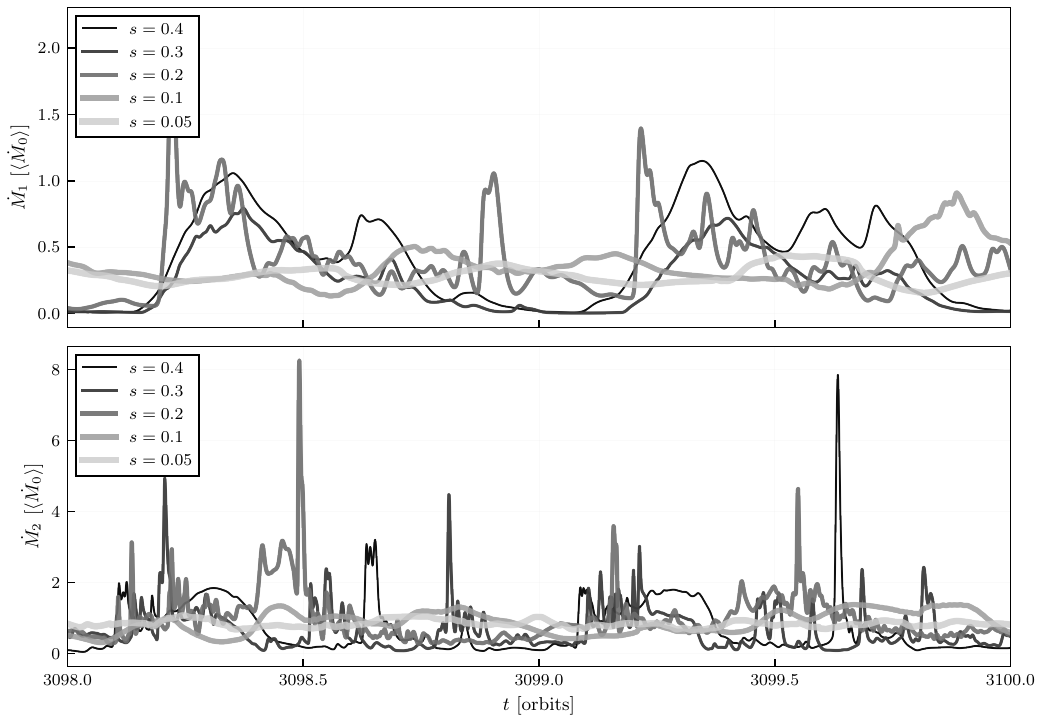}
  \caption{Raw accretion rate time series $\dot M_{1,2}(t)$ over a 100-orbit window, for various $s$ values. Variability diminishes as $s$ decreases.}
  \label{fig:raw_accretion_rate}
\end{figure*}

\begin{figure*}[!htbp]
  \centering
  \includegraphics[width=\textwidth]{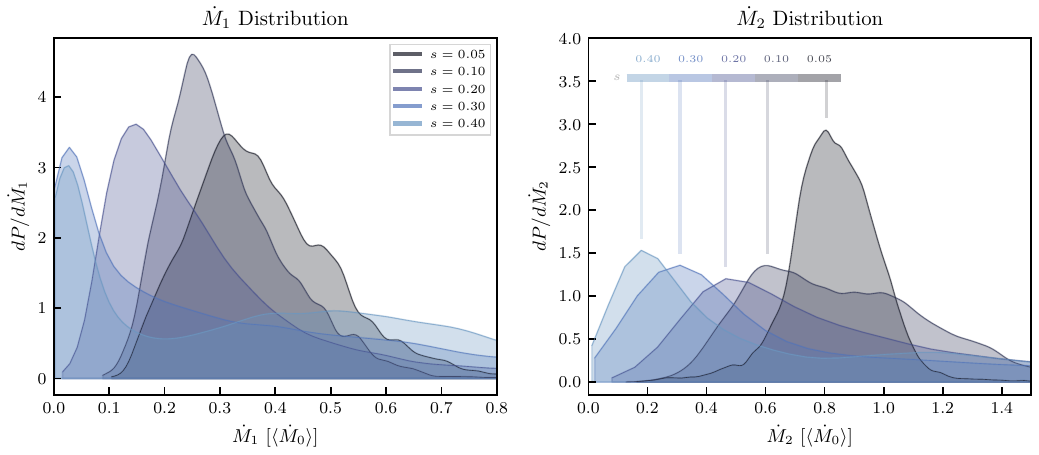}
  \caption{Probability density distributions of $\dot{M}_1$ (top) and $\dot{M}_2$ (bottom) for different $s$ values, with $\dot{M}$ in units of $\langle\dot{M}_0\rangle$. Distributions narrow systematically as $s$ decreases.}
  \label{fig:mdot_pdfs}
\end{figure*}

\begin{figure}[!htbp]
  \centering
  \includegraphics[width=\columnwidth]{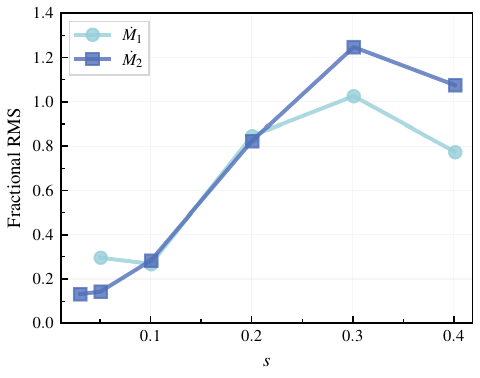}
  \caption{Fractional rms variability amplitude of $\dot{M}_1$ and
$\dot{M}_2$ as a function of $s$.}
  \label{fig:rms}
\end{figure}

Figs.~\ref{fig:raw_accretion_rate} and \ref{fig:mdot_pdfs} show the same behavior in the time and amplitude domains, with raw time series becoming less spiky and accretion rate probability distribution functions narrowing as $s$ decreases. Fig.~\ref{fig:rms} shows that the fractional rms amplitude decreases with $s$ and approaches $\simeq 0.15$ by $s = 0.05$. This flattening suggests the minidisk acts increasingly as an ideal low-pass filter in the small-sink limit. The saturation of the high-frequency PSD also means that the fractional rms as a function of $s$ (Fig.~\ref{fig:rms}) shows a horizontal asymptote at small $s$. This is also because the PSD is steeper than $-2$, so the total variation converges.

\subsection{Radial origin of the variability}
\label{sec:minidisk_inflow}

To identify where the sink variability is generated, we measure the net mass flux time series through a circular control surface of radius $r_{\rm cs}$ centered on $M_2$, spanning $r_{\rm cs}/a \approx 0.008$ (at the sink) to $r_{\rm cs}/a \approx 0.22$ (outside the minidisk's outer edge). The integral covers the full azimuth and includes both inward and outward contributions.

\begin{figure*}[t]
  \centering
  \includegraphics[width=\textwidth]{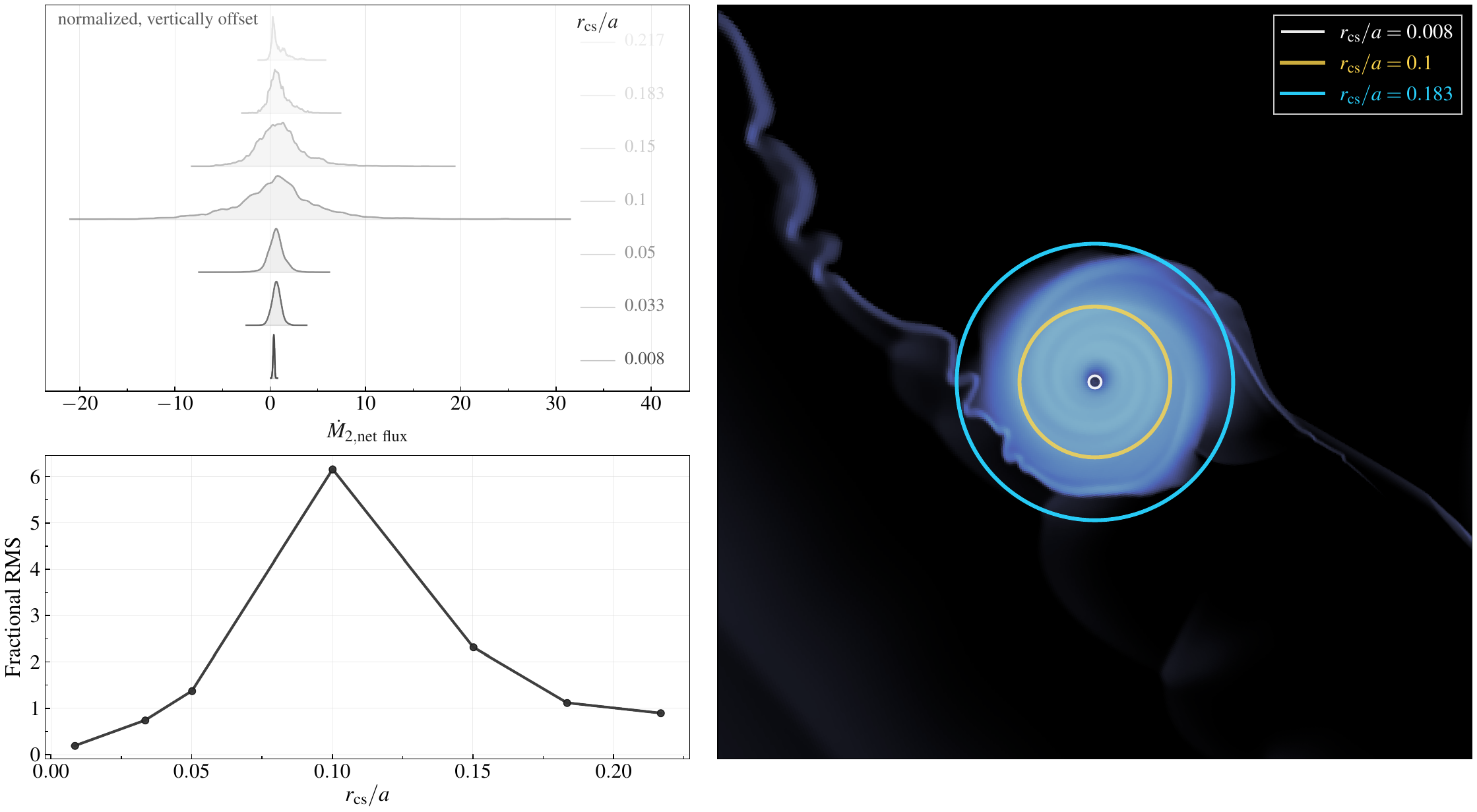}
  \caption{Fractional rms of the net mass flux through circular control
surfaces centered on the secondary, as a function of $r_{\rm cs}/a$ for
$s=0.05$. The profile peaks at $r_{\rm cs}/a\approx0.10$, in the
circularization/sloshing region, and declines on both sides.}
\label{fig:minidisk_inflow_frms}
\end{figure*}

\begin{figure}
  \centering
  \includegraphics{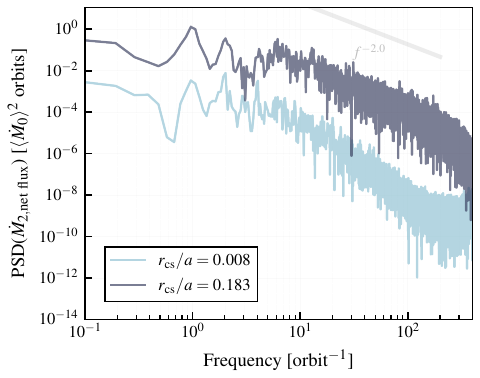}
  \caption{Smoothed PSDs of the net control-surface mass flux centered on $M_2$ for $s = 0.05$, at two representative radii. The grey reference line is an $f^{-2}$ slope.}
  \label{fig:minidisk_inflow_psds}
\end{figure}

Fig.~\ref{fig:minidisk_inflow_frms} shows a non-monotonic radial profile in the fractional rms of the net mass flux. The profile peaks at $r_{\rm cs}/a \approx 0.10$ and declines on both sides. The peak is produced by two effects acting together: large absolute sloshing amplitudes, and partial cancellation between comparable inward and outward gross fluxes that suppresses the mean net throughput. Outward from the peak, the rms declines as the contour enters the quasi-steady circumbinary flow. Inward of the peak, the fractional rms declines as gas settles into approximately Keplerian motion and the mean throughput recovers.

We identify the outer edge of the secondary minidisk with $r \simeq 0.18$--$0.20\,a$, where Fig.~\ref{fig:minidisk_inflow_frms} transitions from the disturbed inner profile to the quasi-steady circumbinary flow. The control-surface PSDs in Fig.~\ref{fig:minidisk_inflow_psds} show how the spectral shape depends on the measurement radius. At $r_{\rm cs}/a = 0.008$, inside the circularized minidisk, the PSD is steep at high frequencies, consistent with minidisk filtering of the incoming supply. At $r_{\rm cs}/a = 0.183$, near the outer minidisk and stream-feeding interface, the PSD is shallower, and approaches the $f^{-2}$ reference slope, consistent with an Ornstein-Uhlenbeck (OU) supply driver. In Fig.~\ref{fig:minidisk_inflow_psds} the outer surface has the higher normalization, since it sits near the stream-fed outer edge where the supply is noisiest, and the minidisk smooths the variability as the gas moves inward.

\subsection{Driver-filter stochastic process}
\label{sec:cutoff}

We decompose the sink PSD as
\begin{equation}
P_{\rm sink}(f) = P_{\rm drv}(f)\,\left|H(f)\right|^2 ,
\label{eq:psd_factor}
\end{equation}
where $P_{\rm drv}$ is the driver PSD supplied at the outer edge of the minidisk and $|H|^2$ is the squared transfer function of the minidisk filter. In the resolved band the supply contributes a red-noise continuum, and above the outer-minidisk break $f_d$ the filter adds a further $f^{-2}$, giving the steep high-frequency tail. The analytic model is developed in Sec.~\ref{sec:minidisk_analytic}.

The break $f_d$ is the local Keplerian frequency at the outer edge of the secondary minidisk,
\begin{equation}
f_d \approx f_{K,2}(r_{\rm out}).
\end{equation}
The control-surface diagnostics place this radius at $r_{\rm out} \simeq 0.18$--$0.20\,a$. For $q = 0.2$, $f_{K,2}(r_{\rm out}) \simeq 5$, consistent with the observed break. This radius is set by binary geometry and does not depend on $s$.

For large sinks the upper break tracks the local Keplerian frequency at the imposed sink radius,
\begin{equation}
f_{{\rm cut},i} \approx f_{K,i}(R_i) = \frac{M_{\rm bin}}{M_i}\,s^{-3/2},
\label{eq:fcut_largesink}
\end{equation}
where $M_i$ is the local component mass. The cutoff is therefore component-specific: because $M_2 < M_1$, the secondary cutoff lies at higher frequency than the primary at fixed $s$. This regime applies to our $s \geq 0.2$ runs.

For small sinks, decreasing $R_i$ no longer moves the cutoff to higher frequency. The sink-controlled break disappears from the measured band, leaving the small-sink continuum whose possible orbital-to-outer-minidisk $f^{-2}$ segment is too narrow to resolve at $q = 0.2$. Thus $f_{\rm cut}$ is a large-sink diagnostic, not a separate physical break in the converged small-sink limit.

The fractional rms peak in Fig.~\ref{fig:minidisk_inflow_frms} at $r_{\rm cs}/a \approx 0.10$ sits at a smaller radius than $r_{\rm out}$. This is expected, because the two diagnostics measure different quantities. The fractional rms peaks in the sloshing zone where the mean net flux is most suppressed by sign cancellation, while $f_d$ is the dynamical frequency at the outer minidisk edge, the radius that sets the onset of filtering.

\subsection{Lack of sensitivity to viscosity}
\label{sec:viscosity_sensitivity}

\begin{figure}
  \centering
  \includegraphics{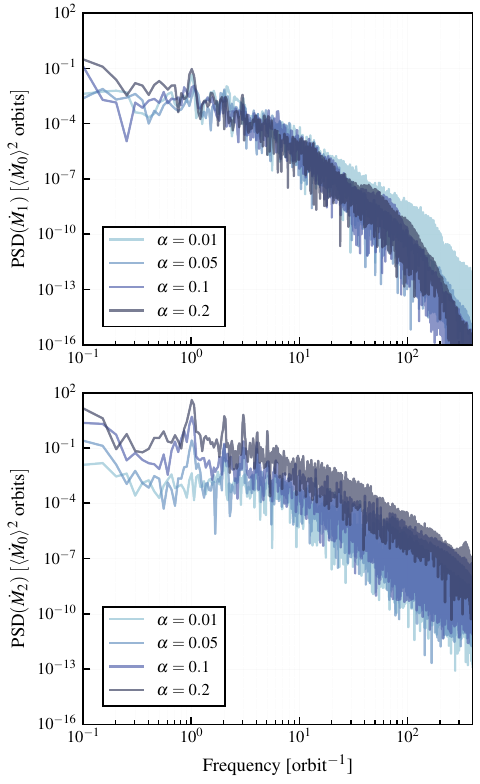}
  \caption{Power spectral densities of $\dot M$ at $s = 0.05$ for $\alpha \in \{0.01,\,0.05,\,0.1,\,0.2\}$.}
  \label{fig:alphapsds}
\end{figure}

At $s = 0.05$, where the sink-controlled break is no longer separately resolved, the PSD shape is nearly unchanged across $\alpha \in \{0.01,\,0.05,\,0.1,\,0.2\}$ (Fig.~\ref{fig:alphapsds}). Both the break position and the $-4$ high-frequency slope are insensitive to the viscosity over an order of magnitude. The low-frequency plateau amplitude, however, does depend on $\alpha$. For the secondary it increases by roughly three orders of magnitude from $\alpha = 0.01$ to $0.2$. An $\alpha$-dependent filter would shift the break or slope instead. We therefore attribute the amplitude dependence to the driver, which injects more variable stream feeding at the minidisk edge, raising $P_{\rm drv}(0)$ without changing its shape.

A viscous-diffusion timescale $t_{\rm visc} \sim (\Delta r)^2/\nu$ would scale as $\nu^{-1} \propto \alpha^{-1}$, so the corresponding break frequency would scale as $\alpha$ and shift by a factor of twenty across $\alpha = 0.01$--$0.2$. No such shift is observed. The absence of any $\alpha$ dependence is therefore consistent with the small-sink break being a dynamical frequency set by minidisk extent rather than a viscous diffusion frequency. Viscosity plays only an indirect role, enabling the minidisk to form and setting its steady-state mass, while the break location in the resolved small-sink limit is dynamical.

\section{Analytic model for minidisk filtering}
\label{sec:minidisk_analytic}

Sec.~\ref{sec:cutoff} interprets each component minidisk as a buffer that converts an intermittent mass supply from circumbinary streams into a smoother accretion rate through the sink. Because the minidisk has a finite dynamical response time, it acts as a low-pass filter. Slow variations in the supply are transmitted with little attenuation, while rapid fluctuations are damped before reaching the inner boundary. The relevant response time is dynamical, set at the minidisk outer edge, rather than the much longer viscous draining time (Sec.~\ref{sec:viscosity_sensitivity}). In the Fourier domain, the sink and supply PSDs are therefore related by
\begin{equation}
  P_{\rm sink}(f) = |H(f)|^2 \, P_{\rm drv}(f),
  \label{eq:PSD_map}
\end{equation}
where $P_{\rm drv}$ is the driver PSD injected at the minidisk outer edge, and $|H(f)|^2$ is the squared modulus of the minidisk transfer function. This factorization separates the origin of variability into an external driver $P_{\rm drv}$, governed by the circumbinary feeding pattern, and an internal response $|H|^2$, governed by transport through the minidisk.

\subsection{PSD template}
\label{sec:results_template}

Sec.~\ref{sec:minidisk_inflow} showed that the control-surface PSD at large $r_{\rm cs}/a$ has a flat-to-$f^{-2}$ morphology consistent with an Ornstein--Uhlenbeck process. As an effective description of the spectrum supplied to the inner minidisk in the resolved band, we model it with a single Lorentzian break at the outer-minidisk frequency $f_d$,
\begin{equation}
  P_{\rm drv}(f) = \frac{A}{1 + (f/f_d)^2},
  \label{eq:p_drv}
\end{equation}
with $f_d \approx f_{K,2}(r_{\rm out}) \simeq 5$ for $q = 0.2$. This effective form should not be interpreted as a measurement of the lowest driver break. An additional lower continuum break near the binary orbital frequency is possible, as discussed in Sec.~\ref{sec:doubly_broken}, but the present $q = 0.2$ runs do not cleanly resolve it. Modelling the minidisk as a single dominant low-pass stage with break $f_{\rm cut}$,
\begin{equation}
  |H(f)|^2 = \frac{1}{1 + (f/f_{\rm cut})^2},
  \label{eq:H_square}
\end{equation}
and combining with Equation~\eqref{eq:p_drv} gives the sink PSD template,
\begin{equation}
  P(f) = \frac{P_0}{\left[1 + (f/f_d)^2\right]\left[1 + (f/f_{\rm cut})^2\right]}.
  \label{eq:psd_template}
\end{equation}

Equation~\eqref{eq:psd_template} is the simplest closed-form expression consistent with the observed PSD morphology. It is flat at low frequencies, has slope $-2$ between $f_d$ and $f_{\rm cut}$, and slope $-4$ above both breaks. When the two breaks merge ($f_{\rm cut} \to f_d$), as in the small-$s$ limit, the template reduces to the Lorentzian-squared form
\begin{equation}
  P(f) = \frac{P_0}{\left[1 + (f/f_b)^2\right]^2}, \qquad f_b \equiv f_d \approx f_{\rm cut},
  \label{eq:psd_template_limit}
\end{equation}
with asymptotic slope $\beta = 4$ at $f \gg f_b$. Fitting Equation~\eqref{eq:psd_template_limit} to the $s = 0.05$ PSD yields $f_b \simeq 5$, consistent with $f_{K,2}(r_{\rm out})$.

The $-4$ slope predicted by this model is steeper than the $f^{-2}$ expectation of damped-random-walk prescriptions commonly applied to single-AGN variability \citep{Kelly2009}. The model therefore predicts that binary AGN candidates with buffering minidisks should exhibit high-frequency slopes approaching $-4$, rather than the $-2$ of a simple DRW.

\subsection{Comparison to the propagating-fluctuations model}
\label{sec:propfluc_comparison}

In the propagating-fluctuations model \citep{Lyubarskii1997}, stochastic stress fluctuations are generated throughout the disk and propagate inward, producing broadband red noise whose slope is set by the assumed statistics of the local driving. The minidisk filtering picture differs in one important respect, in that the dominant forcing is external to the minidisk body. The stream-fed inflow at the minidisk outer edge provides $P_{\rm drv}(f)$, and the minidisk reshapes that signal through the transfer function $|H(f)|^2$ rather than through an integral over locally generated noise sources at all radii. In this picture the PSD is controlled by Equation~\eqref{eq:PSD_map}, and the high-frequency slope is set by the filter order rather than by a distributed stochastic driver.

\subsection{Connection to the two regimes in the simulations}
\label{sec:regimes_connection}

The break frequencies in Equation~\eqref{eq:psd_template} are set by local dynamical times. The lower break in the effective resolved-band template is the outer-minidisk break, $f_d \approx f_{K,2}(r_{\rm out})$. For shallow minidisks associated with large sinks ($s \gtrsim 0.2$), the upper break tracks the imposed sink radius, $f_{\rm cut} \sim f_{K,i}(R_i) \propto s^{-3/2}$, so the two breaks separate and the template formally contains an intermediate $f^{-2}$ segment between them. In the runs this segment is not cleanly resolved, and the large-sink PSD appears as a broadened high-frequency turnover rather than a distinct three-segment power law. For deep minidisks associated with small sinks ($s \lesssim 0.1$), $f_{\rm cut}$ stops tracking the sink radius and is no longer separately resolved, so Equation~\eqref{eq:psd_template} reduces to the Lorentzian-squared form of Equation~\eqref{eq:psd_template_limit}. This reduction describes the measured $q = 0.2$ spectra, while the possible lower orbital break of the physical small-sink continuum is discussed in Sec.~\ref{sec:doubly_broken}.

\section{Discussion}
\label{sec:discussion}

The PSD model of Sec.~\ref{sec:results} and \ref{sec:minidisk_analytic} connects each component minidisk to an external driver and a dynamical filter response. We now relate these results to AGN variability, outline their observational signatures, and list the main limitations.

\subsection{Similarity with single black hole AGN PSDs}
\label{sec:agn_comparison}

Observed AGN PSDs frequently exhibit high-frequency slopes steeper than the $-2$ expected from a damped random walk. \citet{Smith2018} analyzed PSDs for 21 Type-1 AGN observed with Kepler, finding several sources with slopes in the range $-2.5$ to $-3.3$. \citet{Mushotzky2011} reported slopes between $-2.6$ and $-3.3$ for a smaller high-cadence sample. The Kepler/K2 sample of \citet{Aranzana2018} spans $-2$ to $-3$, and \citet{Arevalo2024} extends this to slopes as steep as $-4$. These slopes exceed the DRW prediction \citep{Kelly2009,MacLeod2010}.

Our binary simulations produce an asymptotic $-4$ slope above both breaks of Equation~\eqref{eq:psd_template}. If a subset of the AGN with the steepest observed slopes harbor sub-parsec binaries with buffering minidisks, then minidisk filtering offers a natural explanation for those slopes. This is a falsifiable prediction. Large PSD surveys \citep{Burke2021} can test whether the steepest slopes occur preferentially in systems with independent binary indicators.

The usual explanation for AGN red noise is the propagating-fluctuation model \citep{Lyubarskii1997}, which gives flicker noise with a spectral index near $-1$, shallower than what is typically seen. Our binary PSDs run steeper, near $-4$. A follow-up study with a realistic cooling law and self-consistent emission will look at the optical PSDs directly, which we expect could show more of the $f^{-2}$ noise from the streams, produced where the tidal streams shock against the minidisks. Radiation MHD simulations of both single and binary accretion could help separate the two pictures.

The same physics points to a tentative signpost for individual systems. If an AGN is already suspected of hosting a binary, from a periodicity candidate or a multi-epoch shift in its broad-line velocity, our model predicts what its continuum should look like: a steep slope approaching $-4$, with a break a few times above the suspected orbital frequency. Because that break ratio is fixed by Roche-lobe geometry and barely depends on mass ratio (Sec.~\ref{sec:doubly_broken}), the predicted break near $5\,f_{\rm bin}$ does not require knowing the secondary mass. Finding the break there, with the steep slope, would support a binary reading of a candidate found by other means. Not finding it would weigh against one.

On the observational side, large PSD surveys can test whether the steepest slopes show up preferentially in AGN with independent signs of a binary, such as multi-epoch broad-line velocity drifts \citep{Shen2010,Guo2019} or well-vetted periodic candidates. A binary with a buffering minidisk should show a sharp cutoff near $f_d \simeq 5$ in binary-orbit units, with little power above it. Reverberation mapping, which traces response-function delays, could further separate disk-intrinsic from binary-induced variability \citep{Fu2025}.

\subsection{Should the PSD be a doubly broken power law?}
\label{sec:doubly_broken}

Our driver-filter model involves two break frequencies (Equation~\eqref{eq:psd_template}), so doubly broken power laws might in principle be expected, with the disparity between the breaks depending on the binary mass ratio. We find instead that if the two breaks are the binary orbital frequency and the Keplerian frequency at the secondary minidisk outer edge, their ratio is only weakly sensitive to $q$. The Roche-lobe geometry ties the orbital period at the disk edge to the binary period, so the intermediate $f^{-2}$ segment is generically narrower than one decade. We therefore predict that sink accretion rate PSDs generically appear as broken power laws with a single broad knee, rather than as two well-separated breaks.

The weak $q$-sensitivity follows from the Roche-lobe scaling of the minidisk outer edge. Writing $r_{\rm out}/a \propto q^{\eta}$ and using $M_2/M_{\rm bin} \simeq q$ for $q \ll 1$, the break ratio scales as
\begin{equation}
  \frac{f_d}{f_{\rm bin}} \sim f_{K,2}(r_{\rm out}) \propto q^{(1-3\eta)/2},
  \label{eq:fd_scaling}
\end{equation}
so that $q \propto (f_d/f_{\rm bin})^{2/(1-3\eta)}$, where $f_{\rm bin}$ is the binary orbital frequency. If the outer edge tracks a fixed fraction of the secondary Roche lobe, the Eggleton approximation \citep{Eggleton1983},
\begin{equation}
  \frac{R_{{\rm L},2}}{a} = \frac{0.49\,q^{2/3}}{0.6\,q^{2/3} + \ln(1 + q^{1/3})},
  \label{eq:eggleton}
\end{equation}
gives $r_{\rm out} \propto a\,q^{1/3}$, hence $\eta = 1/3$. The exponent $(1-3\eta)/2$ then vanishes, and $f_d/f_{\rm bin}$ is independent of $q$ to leading order, with only a weak residual from the $M_2/M_{\rm bin} = q/(1+q)$ factor. The two physical break frequencies therefore do not provide a simple estimator of the form $q = (f_{\rm break,1}/f_{\rm break,2})^n$.

This approximate $q$-invariance of the break spacing is itself a testable prediction. A measured dependence of $f_d/f_{\rm bin}$ on $q$ would imply that $r_{\rm out}$ departs from Roche-lobe scaling, and would directly measure the exponent $\eta$ in the minidisk-size relation. The mass-ratio information should instead enter through the visibility and amplitude of the $f^{-2}$ segment. Existing simulations show sharp changes in accretion morphology with $q$, including the transition to a lopsided, strongly fluctuating cavity near $q \simeq 0.04$ \citep{Dorazio2016} and highly variable accretion above $q \simeq 0.05$, with significant four-to-five-orbit variability appearing above $q \simeq 0.2$ \citep{Duffell2020}. Such lump and orbital-frequency features can overlap the narrow $f^{-2}$ segment and obscure it. Lower-$q$ simulations may therefore reveal the segment more clearly by reducing periodic contamination, even if the intrinsic separation between the breaks does not increase. This is a hypothesis to be tested with a dedicated $q$ survey at fixed sink prescription, Mach number, and viscosity.

The lower break is not something we impose by hand. It sits at the orbital frequency because the streams deliver gas to each minidisk once per binary orbit, so the supply itself carries that cadence. The supply can also carry lower-frequency structure associated with the eccentric cavity or ``lump'' near $f \approx 0.15$ \citep{MacFadyenMilosavljevic2008}, so $f = 1$ is the natural orbital feeding cadence rather than a strict lower bound on supply variability. The existence and location of the lower break should be tested with longer control-surface time series and with simulations at other mass ratios. Observed optical power spectra already show more than one continuum scale, and \citet{YukDai2025} identify a second, higher-frequency break in optical AGN PSDs using combined ASAS-SN and TESS light curves. These breaks need not have a binary origin, but they show that doubly broken optical PSDs are measurable targets for timing analyses.

\subsection{Are AGNs stochastically variable because they are all binaries?}
\label{sec:binary_scope}

A main result of this study is that binary-disk interaction generically leads to accretion rate time series with broken power-law PSDs, resembling the canonical noise spectrum observed across the AGN population. Indeed, many AGNs exhibit optical PSDs that transition from flat to red noise, with slopes between $-2$ and $-4$, at break frequencies corresponding to timescales of roughly 100 to 1000 days \citep{Burke2021}. Our results thus seem to justify a serious evaluation of the conjecture that the stochastic variability of AGN indicates that secondary massive black holes are ubiquitous in the AGN population.

To conjecture that such AGN host binary SMBHs implies that binaries are extremely common, and that binarity and fueling episodes arise together, perhaps with a common cause in the galaxy merger that formed the binary \citep[e.g.][]{MihosHernquist1996}. A relevant consideration is the inspiral regime of the implied systems. At the masses that pulsar timing constrains, $\gtrsim 10^{8}\,M_\odot$, an orbital period of $500$ to $5000$ days places the binary well inside the regime where gravitational radiation drives the orbital decay: the merger time (\citep{Peters1964}) is $\sim 10^{4}$ to $10^{7}$\,yr, far shorter than the gas-driven inspiral time. Gas torques, with $\dot a / a \sim -\dot M / M$ \citep{Haiman2009,Munoz2019,Duffell2020,Zrake2021,Lai2023,Clyburn2025}, set the decay only at longer periods, of order decades to a century, where the residence time lengthens to the Salpeter time, $4 \times 10^{7}$\,yr \citep{Salpeter1964}, comparable to the duration of an AGN episode \citep[e.g.][]{Shlosman1990}. The systems whose turnovers fall at $100$ to $1000$ days are therefore gravitational-wave driven, which is exactly why they fall in the band that pulsar timing measures.

However, the amplitude of the $1$--$100$\,nHz stochastic gravitational-wave background is constraining \citep{NANOGrav2023}. In our model the break sits a few times above the binary orbital frequency, $f_d \simeq 5\,f_{\rm bin}$, so a turnover at $100$ to $1000$ days implies an orbital period of roughly $500$ to $5000$ days. The corresponding gravitational-wave frequency, $f_{\rm GW} = 2/P_{\rm bin}$, is then about $5$ to $50$\,nHz, within the pulsar-timing band. If a large fraction of luminous AGN were binaries at these periods, together they would overproduce the background, which already requires the binaries that dominate it to be a minority of massive systems above $10^8\,M_\odot$ at redshifts below about $2.5$ \citep{CaseyClyde2022}. Pulsar timing experiments may therefore exclude the possibility that nearly all AGN host such binaries, provided the measured turnovers are intrinsic to the sources. A turnover timescale is reliable only when the light curve is at least ten times longer than the timescale itself \citep{Kozlowski2017}, so a reported timescale approaching the survey baseline should be treated with caution. It’s noteworthy that the canonical AGN variability being explained by widespread binarity seems to be uniquely excluded by the results of pulsar timing experiments, it exemplifies a manner in which low-frequency GW observations may be used to constrain astrophysical processes.  

\subsection{Implications for periodicity searches}
\label{sec:periodicity}

In the small-sink limit, the minidisk suppresses stochastic power above the outer-minidisk break, $f_d \simeq 5$ in binary-orbit units. Close binaries could therefore display orbital modulation against a reduced high-frequency stochastic background. Many candidate sub-parsec binaries have been identified through periodic or quasi-periodic optical light curves \citep{Charisi2016,Charisi2022,Chen2024}, though distinguishing genuine periodicity from stochastic red noise remains difficult \citep{Vaughan2016,Witt2022}.

Our results suggest that a deficit of high-frequency power can itself serve as a binary indicator. Detecting this morphology does not require observing a full orbit. In q = 0.2 simulations, the continuum break lies at $f_d \simeq 5$ in binary-orbit units, so the diagnostic feature sits at a timescale several times shorter than the orbital period, and fast-cadence monitoring over roughly one binary orbit is sufficient to resolve it.

\subsection{Connection to X-ray fractional variability}
\label{sec:fvar_connection}

X-ray timing studies usually quote variability as the fractional rms $F_{\rm var}$, the scatter of the light curve divided by its mean, after removing measurement noise \citep{Vaughan2003}. We can compute the same quantity from our simulations, using the mass flux through circles of different radius around the secondary (Fig.~\ref{fig:minidisk_inflow_frms}).

The result depends strongly on which radius you look at. The flux varies most, by $50$ to $70$ percent, in the region where the stream circularizes and sloshes around the minidisk. It varies less inside the minidisk, about $30$ percent, and least of all by the time it reaches the sink, about $15$ percent (Fig.~\ref{fig:rms}). The gas gets smoother as it moves inward. So if the X-ray emission comes mostly from the inner disk, it should show weaker fast variability than an optical signal tied to the noisier outer region where the streams hit. How much variability you measure also depends on the length of the light curve relative to where the PSD bends \citep{Vaughan2003,McHardy2006}. This difference between the outer and inner gas is a signature of circumbinary accretion that photon frequency-resolved rms measurements could pick up.

\subsection{Numerical implications}
\label{sec:numerical_implications}

The small-sink limit does two things for a simulation. It resolves the full radial extent of each minidisk, and it lets the minidisk smooth the fluctuating stream supply before the gas reaches the sink. A large sink ($s \gtrsim 0.2$) cuts the minidisk off from the inside and lets extra high-frequency power through, which shows up in the accretion rate as large spikes. When the black holes are meant to be unresolved on the grid, those spikes are an artifact of the oversized sink, not a real feature. Once the sink is small enough, below $s \simeq 0.1$, shrinking it further stops changing the variability that reaches the sink (Fig.~\ref{fig:rms}), because the sink is no longer what sets the size of the minidisk. Studies that use larger sinks should check that the PSD and the variability amplitude have stopped changing with sink size before treating large accretion-rate spikes as physical.

\subsection{Limitations and future directions}
\label{sec:limitations}

The present study fixes several parameters. We have used one mass ratio ($q = 0.2$), one Mach number ($\mathcal{M} = 10$), and a circular orbit. We have shown (Sec.~\ref{sec:viscosity_sensitivity}) that the PSD is not sensitive to the viscosity coefficient $\alpha$ across an order of magnitude, and we suspect it is also not sensitive to the form of the viscosity prescription, though we have not checked this. The sink size at which the small-sink limit sets in, the value of $f_d$, and how $f_{\rm cut}$ depends on sink size may all change with mass ratio or disk thickness. An eccentric binary would feed the streams on additional timescales \citep{Siwek2023,Lai2023}. Mapping out this parameter space could be fruitful.

A more serious limitation of the present study is the locally isothermal equation of state, which sets the gas temperature from the gravitational potential instead of evolving it with an energy equation. The temperature does vary in space and time, since the potential does, but it cannot respond to local heating or cooling. The only time-dependent observable is $\dot M(t)$, which we use as a stand-in for the bolometric luminosity. Replacing the isothermal assumption with an energy equation that includes viscous heating and radiative cooling would give a self-consistent disk surface temperature $T_{\rm eff}(r, \phi, t)$. From that one could evaluate the Planck function $B_\nu(T_{\rm eff})$ and integrate over the disk in any band to get synthetic light curves $L_\nu(t)$. For black hole masses of $10^6$ to $10^8\,M_\odot$, the inner minidisk reaches temperatures that peak in the extreme UV, extending into the soft X-ray for lower-mass and higher-Eddington systems, while the outer minidisk contributes in the UV and optical \citep{Done2012,Cai2023}. Because different bands come from different radii, the filtering should be wavelength-dependent: $f_d$, $f_{\rm cut}$, and the degree of high-frequency suppression would shift from band to band, a fingerprint that a single $\dot M$ proxy cannot capture. These simulations are still tractable in the 2D $\alpha$-viscosity framework used here, and GRMHD simulations could test whether magnetic stresses or radiation pressure change the picture, though at a cost that currently limits how long they can run.

A further advantage of the isothermal approach is the run length it affords. Mapping the continuum PSD means following the accretion rate at high cadence over many binary orbits, which ours does for 100. Radiative magnetohydrodynamic calculations, which evolve the disk temperature self-consistently, are far more costly and reach shorter baselines. The first radiative minidisk run \citep{Chan2025} spent millions of core-hours on a few binary orbits, only a fraction of one in full radiative magnetohydrodynamics, and a global circumbinary run reaches only tens of orbits at similar expense \citep{Tiwari2025}. The turnover we propose as a binary signature sits a few times above the orbital frequency, so it is the part of the spectrum most accessible to short records and fast cadence, in simulations and observations alike. The low-frequency plateau and the orbital break are what demand a long baseline, where an affordable isothermal run is for now the practical route, and these PSDs give future radiative work a target to reproduce.

\section{Summary}
\label{sec:summary}

We have measured the power spectral density (PSD) of the time series of component accretion rates in an unequal-mass ($q=0.2$) circular binary, using very high resolution hydrodynamics simulations in a thin-disk approximation with locally isothermal equation of state, systematically exploring the role of a (potentially) numerical sink size parameter $s$. Our work is motivated by the hope to reveal a signature in the \emph{continuum} PSD of binary-hosting AGN emission, to complement already well-studied \emph{periodicity} signatures. Such signposts showing up in the continuum PSD could in principle facilitate the identification of new binary candidates with long orbital periods, or help corroborate existing candidates.

We find that the power spectrum of the accretion is a broken power law, being flat at low frequencies, and steepening to a slope of $-4$ above a break a few times higher than the binary orbital frequency.

We show that this universal PSD emerges from a simple picture, in which the circumbinary streams feed the minidisks as a white-noise process cutting off  near the orbital frequency, and the minidisks act as low-pass filters acting at the Keplerian frequencies of their outer edges. This picture predicts a double-broken power law, formally exhibiting a short $-2$ slope around frequency $f \simeq 1 - 5$ in units of the inverse binary orbital period. In practice our computed PSDs show a ``broad knee'' around $f \simeq 1 - 5$. The universal PSD is realized for small sinks, $s \lesssim 0.05$. Larger sinks produce a high-frequency excess associated with accretion rate spikes that should be seen as artificial when the sink size is larger than the size of the black hole innermost stable circular orbit.

Our PSD resembles the broadband stochastic variability seen across the AGN population, which invites the question of whether secondary black holes are widespread in active nuclei. However the canonical AGN noise spectrum commonly exhibits a break around $\unit[10^{2-3}]{days}$, which according to our results would put the orbital periods in the year-to-decade range, squarely in the sensitivity range of pulsar timing experiments. It has already been established \citep[e.g.][]{CaseyClyde2022} that if most AGN were year-to-decade binaries, they would overproduce the stochastic gravitational-wave background.

The steep high-frequency slope, together with a break near five times the orbital frequency of an independently suspected binary, may serve as a signpost to look for in high-cadence optical surveys. Follow-up work will compute the PSDs of quasi-thermal radiative output in different energy bands.

\begin{acknowledgments}

J.Z. acknowledges support from the National Science Foundation under Grant Number AST-2408034.
\end{acknowledgments}

\begin{contribution}

\end{contribution}

\bibliography{refs}
\bibliographystyle{aasjournalv7}

\end{document}